\documentstyle[11pt]{article}
\begin{document}
\thispagestyle{empty}
\begin{center}

\vspace{1.8cm} {\bf Chiral bosons on Bargmann space associated with A$_r$
Statistics }\\
\vspace{1.5cm} {\bf M. Daoud}$^a${\footnote {\it Facult\'e des
Sciences, D\'epartement de Physique, Agadir, Morocco; email:
m$_-$daoud@hotmail.com}} and {\bf A. Hamama}$^b$
\\\vspace{0.5cm}
$^a$ {\it Max Planck Institute for Physics of Complex Systems,\\ N$\ddot{o}$thnitzer Str. 38, D-01187 Dresden, Germany}\\ \vspace{0.2cm} $^b${\it
High Energy Laboratory, Faculty of Sciences, University Mohamed V,\\
P.O. Box 1014 , Rabat ,
Morocco}\\[1em]
\vspace{3cm} {\bf Abstract}
\end{center}
\baselineskip=18pt
\medskip
We consider a large collection of particles obeying $A_r$
statistics. The system behaves like a quantum droplet
characterized by a constant Husimi distribution. We show that the
excitations of this system live on the boundary of the droplet and
they are described by an effective chiral boson action
generalizing the Wess-Zumino-Witten theory in two dimension. Our
analysis is based on the Fock-Bargmann analytical
 representations associated to $A_r$ statistics.
  The quantization of the theory describing the dynamics on the edge is achieved.
  As by product, we prove that the edge excitations
are given by a tensorial product of $r$ abelian bosonic fields.

\newpage
\section{Introduction}
Fifty years ago, a first extension of Bose and Fermi statistics
was achieved by Green [1]. This extension was the basic underlying
mathematical background to investigate the implications of the
generalizations of the familiar bosonic and fermionic statistics
[2-6]. In the two last decades a renewal interest has been devoted
to generalized quantum statistics due to their possible relevance
in some issues like for instance fractional quantum Hall effect
[7-8], anyon superconductivity [9] as well as black hole
statistics [10]. Many variants of quantum statistics were proposed
in the literature. One may quote the anyonic statistics [4]
interpolating between fermionic and bosonic ones in two dimension
space, quonic statistics [11] developed in the context of
$q$-deformed algebras, $k$-fermionic statistics [12] defined as
$q$-deformed version of ordinary bosons when the deformation
parameter is such that $q^k = 1$  and Haldane
fractional statistics [13] to explain the origin of the fractional quantization of Hall conductivity.\\

In the Green prescription [1], the
generalized quantum statistics are characterized by certain triple relations which
replace the commutation and anti-commutation rules for bosons and fermions. As by product,
the fermions and bosons are promoted to para-fermions and para-bosons of order $r$ respectively where the creation and annihilation operators satisfy
\begin{equation}
\big[[ f_i^+ , f_j^-]  , f_k^- \big] = - 2 \delta_{ik}
f_j^- ,{\hskip 0.5cm} \big[[ f_i^+ , f_j^+]  , f_k^- \big] = - 2
\delta_{ik}
f_j^+ +  2 \delta_{jk}
f_i^+, {\hskip 0.5cm} \big[[ f_i^- , f_j^-]  , f_k^- \big] =
0,
\end{equation}
\begin{equation}
\big[\{ b_i^+ , b_j^-\}   , b_k^- \big] = - 2 \delta_{ik}
b_j^- ,{\hskip 0.5cm}\big[\{ b_i^+ , b_j^+\}   , b_k^- \big] = - 2
\delta_{ik}
b_j^+ - 2 \delta_{jk}
b_i^+,  {\hskip 0.5cm} \big[\{ b_i^- , b_j^-\}  , b_k^- \big] =
0,
\end{equation}
with $i,j,k = 1,2,...,r$ . From an algebraic point of view, the Grassmann algebra in the fermionic case is replaced by
the para-fermionic one (1) which is related to the orthogonal Lie algebra $so(2r+1) = B_r$
[14]. In the other side, the Weyl-Heisenberg algebra is extended to para-bosonic one (2) which
 is connected to the orthosymplectic superalgebra $osp(1/2r) =
B(0,r)$ [15]. This indicates the deep link between Green statistics and
the classical Lie and super Lie algebras. In this vain, very recently, on the basis of Palev works [6 ,
16], a classification of generalized quantum statistics were
derived for the classical Lie algebras $A_r$, $B_r$, $C_r$ and
$D_r$ [17]. \\

This paper concerns the generalized $A_r$ statistics. This
generalization incorporates two kinds of statistics. The first one
deals statistics satisfying a generalized exclusion Pauli
principle and coincides with ones derived by Palev [6-16]. This
class will be termed here fermionic $A_r$ statistics. The second
class is of bosonic kind. The particles can be accommodated in a
given quantum state without any restriction. The generalization of
$A_r$ statistics was derived by one of the authors [18-20].
However,
 for completeness and in order to fix our notations we shall, in the next section,  review the
definition of the generalized $A_r$ by means of the so-called
Jacobson generators [21]. We use the trilinear relations defining
the generalized quantum $A_r$ statistics to construct the
corresponding Fock space. We give the actions of the corresponding
creation and annihilation operators. The spectrum of the
Hamiltonian, describing free particles obeying the generalized
$A_r$ statistics, is determined. The analytical Bargmann
representations corresponding to Fock space of bosonic as well as
fermionic $A_r$ statistics are also presented. We give the
differential realization of Jacobson operators. This analytical
realization provides us with many advantages in discussing the
semiclassical behavior of $A_r$ quantum quantum systems. Indeed,
in the Bargmann space viewed as the phase space of $A_r$ quantum
systems, we study the excitations around a given droplet defined
by a constant density (constant Husimi distribution).
 We show in this paper, that the excitations
live on the boundary of the droplet and are described by an
effective action generalizing Wess-Zumino-Witten one which
describes chiral bosons in two dimensional space time [22]. To
perform the semiclassical analysis of $A_r$ quantum systems and
the derivation of the effective theory of the edges excitations,
some tools are needed like the star product in the Bargmann space.
The necessary material to do this is presented. The strategy that
we adopt is closer to one followed by Das et al [23] and Sakita
[24] for non relativistic fermions localized around a ground many
body state and forming a droplet. The droplet is specified by a
diagonal density matrix $\rho_0$. In the presence of an excitation
potential, the fundamental state can be characterized by a unitary
transformation of $\rho_0$, namely $\rho_0 \longrightarrow U
\rho_0 U^{\dagger}$ where
$U$ is a collective variable describing the possible excitations around the droplet. \\

The paper is organized as follows. In section 2, a brief review of generalized $A_r$ statistics,
 the associated Hamiltonian and the corresponding
Fock space is given. The section 2 also deals with the Bargmann
realization of the Jacobson generators (creation and annihilation
operators of $A_r$ quantum systems). This provides us with an
useful way to analyze semi-classically the system under
consideration. In this sense, we show in section 3 that a large
 collection of particles
obeying $A_r$ statistics behaves like a droplet in the Bargmann space. In this semi-classical description,
 the operators are replaced by functions and the commutators becomes Moyal
brackets. In section 4, we show that the excitations of the system live on the boundary of the droplet and
 they are described by a chiral boson theory. The advantages of such formulation lies on the fact the dynamics
 of the obtained bosonic theory encodes the excitations of the $A_r$ statistical system. It follows that our
 formulation
provides a first convenient step to study large collective states of $A_r$ statistics. The
last section comprises concluding remarks.

\section{The generalized $A_r$ statistics}
In this section, we introduce the definitions of the Jacobson
operators and the generalized $A_r$ statistics viewed as Lie
triple system. We review the construction of the corresponding
Fock space and we give the Hamiltonian describing a quantum system
obeying  generalized $A_r$ statistics [18] (see also [19-20]).
\subsection{Jacobson generators}
First, let us introduce the notion  of Lie triple system. Let $V$ a vector space
over a field $F$ which is assumed to be either real or complex. The vector space $V$
equipped vector  with a trilinear mapping
$$[x,y,z] : V \otimes V \otimes V \longrightarrow V $$ is called Lie
triple system if the following identities are satisfied:
$$[x,x,x] = 0,$$
$$[x,y,z] + [y,z,x] + [z,x,y] = 0,$$ $$[x,y,[u,v,w]] =
[[x,y,u],v,w] + [u,[x,y,v],w] + [u,v,[x,y,w]].$$\nonumber
According this definition, we will  introduce the generalized
$A_r$ statistics as Lie triple system. In this respect, the algebra ${\cal G}$ defined by the generators
 $a_i^+$ and $a_i^- $ ($i = 1, 2,..., r$) mutually commuting $([a_i^- , a_j^-] = [a_i^+ , a_j^+] = 0)$
  and satisfying the triple relation
\begin{equation}
\big[[ a_i^+ , a_j^- ] , x_k^+ \big] = - s  \delta_{jk}
x_i^+ - s  \delta_{ij} x_k^+
\end{equation}
\begin{equation}
\big[[ a_i^+ , a_j^- ] , a_k^- \big] =  s  \delta_{ik}
x_j^- + s  \delta_{ij}  x_k^-,
\end{equation}
where $ s \in \{ 1 , -1 \}$, is
closed under the ternary operation $$[x,y,z] = [[x,y],z]$$
 and
define a Lie triple system. The elements $a_i^{\pm}$ are termed
Jacobson generators and will be identified later with creation
and annihilation operators of a quantum system obeying
generalized $A_r$ statistics. Note that for $ s = - 1$, the
algebra ${\cal G}$ reduces to one defining the $A_r$ statistics
discussed in [6]. As we will see in what
follows, the sign of the parameter $s$ play an importance in
the representation of the algebra ${\cal G}$ and consequently, one
can obtain different microscopic and macroscopic statistical
properties of the quantum system under consideration. Finally, recall that this statistics
is intrinsically related to simple Lie algebras of class $A$ like the para-fermion statistics
which is related to class $B$ of simple Lie algebras.
\subsection{The Hamiltonian}
The Jacobson generators $a_i^{\pm}$ can be identified with
creation and annihilation operators of a quantum gas obeying the
generalized $A_r$ statistics. This requires a consistency with
 the Heisenberg equation
\begin{equation}
[ H , a_i^{\pm} ] = \pm e_i a_i^{\pm}
\end{equation}
where $H$ is the Hamiltonian of the system and the quantities $e_i$ are the energies of the modes
$i = 1, 2, ... , r$. One can verify that if $|E \rangle$ is an
eigenstate with energy $E$, $a_i^{\pm}|E \rangle$ are eigenvectors
of $H$ with energies $E \pm e_i$. In this respect, the operators $
a_i^{\pm}$ can be interpreted as ones creating or annihilating
particles. To solve the consistency equation (5), we write the
Hamiltonian $H$ as
\begin{equation}
H = e_0 {\bf 1} + h =e_0 {\bf 1} + \sum_{i=1}^{r} e_i h_i
\end{equation}
where $e_0$ is an arbitrary real constant and ${\bf 1}$ is an operator
commuting with all the elements of the algebra ${\cal G}$ (belonging to the centre of ${\cal G}$). The Hamiltonian $H$
 seems to be a simple sum of "free" (non-interacting)
Hamiltonians $h_i$. However, it important to note that in the quantum system
under consideration, the statistical interactions occur and are
encoded in the triple commutation relations (3) and (4).
Using the structure relations of the algebra ${\cal G}$, the
solution of the Heisenberg condition (5) is given by
\begin{equation}
h_i = \frac{s}{r+1}\bigg[ (r+1)[ a_i^- , a_i^+ ] - \sum_{j=1}^{r}[ a_j^- ,
a_j^+]
\bigg] + c
\end{equation}
where the constant $c$ will be defined later such that the
ground state (vacuum) of the Hamiltonian $H$ gives the energy $e_0$.
\subsection{Fock representations}
An Hilbertian representation of the algebra ${\cal G}$ can be
simply derived using the relations structures (3-4) defining
$A_r$ statistics. Here, we give the main results. For
more details see the references [18,20]. Since, the algebra ${\cal G}$ is
spanned by $r$ pairs of Jacobson generators, it is natural to
assume that the Fock space ${\cal F}$ is given by
\begin{equation}
{\cal F} =  \oplus_{n=0}^{\infty} {\cal H}^n,
\end{equation}
where $${\cal H}^n
\equiv \{ |n_1, n_2,\cdots , n_r\rangle\ , n_i \in {\bf{N}},
\sum_{i=1}^{r}n_i = n > 0\}$$ and ${\cal H}^0 \equiv {\bf{C}}$.
The action of $a_i^{\pm}$, on ${\cal F}$, are defined by
\begin{equation}
a_i^{\pm} |n_1,\cdots, n_i,\cdots , n_r\rangle\ = \sqrt{F_i
(n_1,\cdots ,n_i \pm 1,\cdots, n_r)}|n_1,\cdots, n_i \pm 1,\cdots
, n_r\rangle\
\end{equation}
where the functions $F_i$ are called the structure functions.
Using the triple structure relations of $A_r$ statistics, one obtains [18,20]
the following expressions
\begin{equation}
F_i (n_1,\cdots ,n_i,\cdots, n_r) = \frac{1}{2}n_i ( 2k - (1+s) + 2s( n_1 +
n_2 + \cdots +  n_r )),
\end{equation}
 in terms of
the quantum numbers $n_1, n_2, \cdots, n_r$. In equation (10), the
real parameter labelling the obtained representation satisfies the
condition
 $
2k-1 > s$. The dimension of the irreducible representation space ${\cal F}$
is determined by the  condition:
\begin{equation}
k - \frac{1+s}{2} + s(n_1+n_2+\cdots+n_r) > 0.
\end{equation}
It depends on the sign of the parameter $s$. It is clear that for
$s=1$, the Fock space ${\cal F}$ is infinite dimensional. However,
for $s=-1$, there exists a finite number of basis states satisfying the
condition $n_1+n_2+\cdots+n_r \leq k-1$. The dimension is given,
in this case, by $\frac{(k-1+r)!}{(k-1)!r!}$. This is exactly the
dimension of the Fock representation of $A_r$ statistics discussed
in [6]. This condition-restriction is closely related to so-called
generalized exclusion Pauli principle according to which no more
than $k-1$ particles can be accommodated in the same quantum
state. In this sense, for $ s = -1$, the generalized $A_r$ quantum
statistics give statistics of fermionic behavior. They will be
termed here as fermionic $A_r$ statistics and ones corresponding
to $s = 1$ will be named bosonic $A_r$ statistics. Having specified the Fock
space associated with the generalized $A_r$ statistics, one can obtain
the spectrum of the Hamiltonian (6). For convenience, we set $c =
\frac{2ks-s-1}{2r+2} $ in (7) and using  the actions of creation and annihilation operators (9-10),
one has
\begin{equation}
H |n_1,\cdots, n_i,\cdots , n_r\rangle\ = (e_0 + \sum_{i=1}^{r}e_i
n_i)|n_1,\cdots, n_i,\cdots , n_r\rangle.
\end{equation}
It is remarkable that, for $s=-1$, the spectrum of $H$ is similar
(with a slight modification) to energy eigenvalues of the $A_r$
Calogero model (see for instance Eq.(1.2) in [25]). The latter
describe the dynamical model containing $r+1$ particles on a line
with long rang interactions and provides a microscopic realization
of fractional statistics [13,26]. It is important to stress that
for $e_i = 0$ for all $i = 1, 2, \cdots, r$, the ground state
energy $e_0$ is degenerate (the degeneracy coincides with the
dimension of the Fock space; It is finite (resp. infinite) for
fermionic (resp. bosonic) $A_r$ statistics). It follows that $h$
in equation (6) can be considered as a potential responsible of
the degeneracy lifting and inducing fluctuations around the ground
state energy $e_0$. This remark constitutes the key ingredient, as
it will be clarified later, to define $A_r$ quantum droplets and
to derive their excitations. Finally, we point out that for large
$k$, we have
\begin{equation}
[ a_i^- , a_j^+] \approx k\delta_{ij}
\end{equation}
reflecting that the generalized $A_r$ statistics (fermionic and
bosonic ones) coincide with the Bose statistics and the Jacobson
operators reduce to Bose ones (creation and annihilation operators
of harmonic oscillators).\\
Besides the Fock representation
discussed in this section, it is interesting to look for
analytical realizations of the space representation associated
with the Fock representations of the generalized $A_r$ statistics.
These realizations  constitute an useful analytical tool in
connection with variational and path integral methods to describe
the quantum dynamics of the system described by the Hamiltonian
$H$.

\subsection{Bargmann realizations}

First note that the Bargmann realization associated to $A_r$
statistics was derived in [18] (see also [19-20]). Here, we recall
some results needed for our task. This realization uses a suitably
defined Hilbert space of entire analytical functions. The Jacobson
annihilation generators $a_i^{\pm}$ are realized as first order
differential operators with respect to a complex variables $z_i$
\begin{equation}
a_i^- \longrightarrow \frac{\partial}{\partial z_i}.
\end{equation}
The key point of such analytical realization lies on the fact that we represent the Fock
 states $\vert n_1, n_2, \cdots, n_r \rangle$
as power of complex variables $z_1, z_2, \cdots, z_r$:
\begin{equation}
| n_1, n_2 \cdots , n_r\rangle  \longrightarrow C_{ n_1,\cdots
,n_r} z_1^{n_1} z_2^{n_2} \cdots z_r^{n_r}.
\end{equation}
Using the action of the annihilation operators on the Fock space
${\cal F}$ and the correspondences (14) and (15), the coefficients $C_{ n_1,n_2,
\cdots,n_r}$ are obtained as

\begin{equation}
C_{ n_1,n_2,
\cdots,n_r}= \bigg[{\frac{(k-1 + sn)!
}{(k-1)!}}\bigg]^{\frac{s}{2}}\frac{1}{\sqrt{n_1!}\sqrt{n_2!}\cdots \sqrt{n_r!}}
\end{equation}
where $n = n_1 + n_2 + \cdots + n_r$ and $s=1$ (resp.$s=-1$) for bosonic (resp. fermionic) $A_r$
 statistics. Using the equations
(15-16), one can determine the differential action of the
Jacobson creation operators. Indeed, from the actions of the
generators $a_i^+$ on the Fock space and the triple relations (3)
and (4) , one obtains
\begin{equation}
a_i^+ \longrightarrow \frac{1}{2}(2k +s - 1)z_i + s z_i \sum_{j=1}^r z_j\frac{d}{dz_j}.
\end{equation}
The Jacobson generators act as first order linear differential
operators. An  arbitrary vector of the Fock
space ${\cal F}$  $$|\phi \rangle =
\sum_{n_1}\sum_{n_2}\cdots
\sum_{n_r}\phi_{n_1, n_2 \cdots , n_r}| n_1,n_2,
\cdots ,n_r \rangle,$$
 is realized as
\begin{equation}
\phi(z_1, z_2 \cdots , z_r) =
\sum_{n_1}\sum_{n_2}\cdots
\sum_{n_r} \phi_{n_1, n_2 \cdots , n_r} C_{n_1, n_2
\cdots , n_r} z_1^{n_1} z_2^{n_2} \cdots z_r^{n_r}.
\end{equation}
The inner product of two functions $\phi$ and $\phi'$ is defined by
\begin{equation}
\langle\phi'|\phi\rangle = \int \int \cdots \int d^2z_1  d^2z_2
\cdots d^2z_r \Sigma(k;z_1,z_2 \cdots ,z_r)\phi'^{\star}(z_1,z_2
\cdots ,z_r) \phi(z_1,z_2 \cdots ,z_r)
\end{equation}
The computation of the integration measure $\Sigma$, assumed
to be isotropic , can be performed by choosing $|\phi\rangle = |
n_1,n_2, \cdots ,n_r \rangle $ and $|\phi'\rangle = | n'_1,n'_2,
\cdots ,n'_r \rangle $. A direct computation shows that the measure can be cast in the following compact
form
\begin{equation}
\Sigma (\varrho_1,\varrho_2, \cdots ,\varrho_r) =
\pi^{-r}\bigg[ \frac{(k-1)!}{(k-sr +\frac{1}{2}(s-1))!}\bigg]^s[1 +s (\varrho_1^2 + \varrho_2^2+
\cdots + \varrho_r^2)]^{sk - r -\frac{1}{2}(s+1)}
\end{equation}
where $\varrho_i = \vert z_i \vert^2$. One can write the function
$\phi(z_1, z_2 \cdots , z_r)$ as the product of the state
$|\phi\rangle $ with some ket $\vert  \bar z_1, \bar z_2 \cdots ,
\bar z_r \rangle$ labelled by the complex conjugate of the
variables $z_1, z_2, \cdots , z_r$
\begin{equation}
\phi(z_1, z_2, \cdots , z_r)= {\cal N} \langle  \bar z_1, \bar z_2,
\cdots , \bar z_r |\phi \rangle,
\end{equation}
where ${\cal N}$ is a normalization constant to be adjusted later. Taking $|\phi\rangle = | n_1, n_2, \cdots , n_r \rangle$, we
have
\begin{equation}
\langle  \bar z_1, \bar z_2,
\cdots , \bar z_r\vert n_1, n_2, \cdots, n_r\rangle = {\cal N}^{-1} C_{n_1, n_2, \cdots, n_r}z_1^{n_1} z_2^{n_2} \cdots z_r^{n_r}.
\end{equation}
This implies
\begin{equation}
| z_1, z_2, \cdots , z_r \rangle =  {\cal N}^{-1}
\sum_{n_1}\sum_{n_2}\cdots
\sum_{n_r} \bigg[{\frac{(k-1 + sn)!
}{(k-1)!}}\bigg]^{\frac{s}{2}}
\frac{z_1^{n_1}}{\sqrt{n_1!}}\frac{z_2^{n_2}}{\sqrt{n_2!}}\cdots \frac{z_r^{n_r}}{\sqrt{n_r!}}.
\end{equation}
It is important to notice that the expansion (23) converges for
bosonic $A_r$ statistics when $\bar z.z = |z_1|^2+ |z_2|^2+ \cdots
+ |z_r|^2 < 1$. In other words, the complex variables $z_1, z_2,
\cdots ,z_r $ should be in the complex domain  defined by$ \{(z_1,
z_2, \cdots, z_r ): |z_1|^2 + |z_2|^2 + \cdots + |z_r|^2 < 1 \}$ .
The normalization constant in (23) is given by
\begin{equation}
{\cal N} = (1 - s(|z_1|^2+ |z_2|^2+ \cdots + |z_r|^2))^{-\frac{1}{4}(2ks - s + 1)}
\end{equation}
The states (23) are continuous in the labeling, constitute an over
complete set in the respect to the measure given by (20) and then
are coherent in the Klauder-Perelomov sense.

\section{Semi classical analysis}
One of the usefulness of the above Bargmann realizations is they provide us with a simply way
to establish a correspondence between operators and classical functions on the phase space of
the systems under consideration. So, in this section we shall investigate the semi classical properties of
$A_r$ statistical systems (bosonic as well as fermionic) in the Bargamnn space
for $k$ large.
\subsection{The density matrix and Husimi distribution}
It is commonly accepted that  the exploration of the classical
behavior of any quantum system hinges on whether one can describe
the behavior of the wave-functions in terms of a density matrix.
So, let $N = N_1 + N_2+\cdots +N_k$ the number of quantas of the
system
 where $N_i$ stands for the particle number in the mode
$i$. The corresponding density operator is
\begin{equation}
\rho_0 = \sum_{n_1}^{N_1}\sum_{n_2}^{N_2}\cdots\sum_{n_r}^{N_r}  \vert  n_1, n_2, \cdots, n_r \rangle \langle  n_1, n_2, \cdots, n_r\vert.
\end{equation}
In the Bargmann space,  the mean value of the density matrix is
defined by
\begin{equation}
\rho_0 (\bar{z} , z)= \langle z \vert \rho_0 \vert z \rangle.
\end{equation}
where $z$ stands for the variables $(z_1, z_2, \cdots, z_r)$
labelling the coherent states for $A_r$ statistics systems. The
mean value $\rho_0 (\bar{z} , z)$ is the symbol associated with
the density operator and it can be identified with the Husimi
distribution for $A_r$ quantum systems. As we are concerned with
the situation when $k$ is large, let us investigate the spacial
shape of the mean value of the density operator. For $A_r$ bosonic
statistics, we have
\begin{equation}
\rho_0(\bar z, z) = (1-\bar z.z)^k \sum_{n_1 = 0}^{N_1}\cdots
\sum_{n_r = 0}^{N_r} \frac{(k-1+n_1\cdots+n_k)!}{(k-1)!n_1!\cdots
n_r!}|z_1|^{2n_1}\cdots |z_k|^{2n_r}
\end{equation}
It is easy to see that, for $k$ large, the identity
\begin{equation}
(1-\bar z.z)^{-k} =  \sum_{n_1 = 0}^{\infty}\cdots \sum_{n_r =
0}^{\infty} \frac{(k-1+n_1\cdots+n_k)!}{(k-1)!n_1!\cdots
n_r!}|z_1|^{2n_1}\cdots |z_k|^{2n_r},
\end{equation}
gives
\begin{equation}
(1-\bar z.z)^{k} =  \exp(-k\bar z.z).
\end{equation}
Furthermore, using the relation
\begin{equation}
\sum_{n_1 = 0}^{N_1}\cdots \sum_{n_r = 0}^{N_r}
\frac{(k-1+n_1\cdots+n_r)!}{(k-1)!n_1!\cdots
n_r!}|z_1|^{2n_1}\cdots |z_r|^{2n_r} = \sum_{n =
0}^{N}\frac{(k-1+n)!}{(k-1)!(n)!}(\bar z.z)^{n}
\end{equation}
where $n=n_1+\cdots+n_r$, one can see the term involving the sum
in the expression of $\rho_0$ behaves like
\begin{equation}
\sum_{n=0}^N \frac{(k\bar z.z)^n}{n!}.
\end{equation}
It follows that , for $k$ large, the density can be approximated by
\begin{equation}
\rho_0(\bar z, z) \simeq \exp(-k\bar z.z)\sum_{n=0}^N \frac{(k\bar
z.z)^n}{n!}\simeq \Theta (N - k\bar z.z)
\end{equation}
for a large number $N$ of particles. Clearly, $\rho_0(\bar z, z)$ is a step
function for $k\longrightarrow \infty$ and $N\longrightarrow
\infty$ ($\frac{N}{k}$ fixed).
Similarly, the classical density for $A_r$ fermionic statistics
\begin{equation}
\rho_0(\bar {z}, z) = (1+\bar{z}.z)^{-\frac{k-1}{2}} \sum_{n_1 = 0}^{N_1}\cdots
\sum_{n_r = 0}^{N_r} \frac{(k-1)}{(k-1-n))!n_1!\cdots
n_r!}|z_1|^{2n_1}\cdots |z_r|^{2n_r}
\end{equation}
gives for large $k$ and $N$
\begin{equation}
\rho_0(\bar z, z) \simeq \Theta (N - k\bar {z}.z).
\end{equation}
It corresponds to a droplet configuration with boundary defined by
$k\bar{z}.z = N$ and its radius is proportional to $\sqrt{N}$. The
derivative of this density tends to a $\delta$ function. As we
will  see an interesting outcome of the semi-quantal dynamics
happens when the parameter $k$ tends to infinity.  This indicates
that $k$ plays a crucial role in determining the classical limit
of $A_r$ quantum dynamics and  in deriving  the edge excitations
for $A_r$ fermionic as well as bosonic statistics.

\subsection{The Star product and Moyal bracket}
A second necessary ingredient to perform our semi classical
analysis is the star product. In fact, as we will discuss next,
for $k$ large the mean value of the product of two operators leads
to the Moyal star product . To show this, to every operator $A$
acting on the Fock space ${\cal F}$, we associate the function
\begin{equation}
{\cal A}(\bar z, z) = \langle z | A | z \rangle.
\end{equation}
An associative star product of two functions ${\cal A}(\bar z, z)$
and ${\cal B}(\bar z, z)$ is defined by
\begin{equation}
{\cal A}(\bar z, z)\star {\cal B}(\bar z, z) = \langle z | AB | z
\rangle = \int d\mu(\bar{z'}, z') \langle z | A | z' \rangle\langle z'| B | z\rangle
\end{equation}
where the measure $d\mu(\bar z, z) = d^2z_1d^2z_2\cdots d^2z_r
\Sigma$ is given by equation (20). To compute this star product,
let us exploit the analytical properties of coherent states
defined above.
 Indeed, Using the equations (23) and (24), one can see
that the function defined by
\begin{equation}
{\cal {A}}(\bar z', z) = \frac{\langle z' | A | z \rangle}{\langle z'  | z \rangle}
\end{equation}
satisfy the following holomorphic and anti-holomorphic conditions
\begin{equation}
\frac{\partial}{\partial \bar{z}_i}{\cal {A}}(\bar z', z) = 0 {\hskip 1cm}\frac{\partial}{\partial z'_i}{\cal {A}}(\bar z', z) = 0
\end{equation}
for $i = 1, 2, \cdots, r$ and $z\neq z'$. Consequently, the action of the translation operator
 on the function ${\cal {A}}(\bar z', z)$ gives
\begin{equation}
\exp\bigg(z'.\frac{\partial}{\partial z}\bigg){\cal {A}}(\bar z', z) = {\cal {A}}(\bar z', z+z')
\end{equation}
from which one can see that the function ${\cal {A}}(\bar z, z')$ is given by
\begin{equation}
\exp\bigg(-z.\frac{\partial}{\partial z'}\bigg)\exp\bigg(z'.\frac{\partial}{\partial z}\bigg){\cal {A}}(\bar z, z) =
 \exp\bigg((z'-z).\frac{\partial}{\partial z}\bigg){\cal {A}}(\bar z, z) = {\cal {A}}(\bar z, z')
\end{equation}
in term of the function ${\cal {A}}(\bar z, z)$. Similarly, one obtains
\begin{equation}
\exp\bigg(-\bar z.\frac{\partial}{\partial \bar z'}\bigg)\exp\bigg(\bar z'.\frac{\partial}{\partial \bar z}\bigg){\cal {A}}(\bar z, z) = {\cal {A}}(\bar z', z)
\end{equation}
Equivalently, the equations (40) and (41) can also be cast in the following forms
\begin{equation}
 \exp\bigg((z'-z).\frac{\partial}{\partial z}\bigg){\cal {A}}(\bar z, z) = {\cal {A}}(\bar z, z')
\end{equation}
and
\begin{equation}
 \exp\bigg((\bar z'- \bar z).\frac{\partial}{\partial \bar z}\bigg){\cal {A}}(\bar z, z) = {\cal {A}}(\bar z', z)
\end{equation}
respectively. Combining the equations (36-37) and (42-43), the star product rewrites as
\begin{equation}
{\cal A}(\bar z, z)\star {\cal B}(\bar z, z) =  \int d\mu(\bar{z'}, z') \exp\bigg((z'-z).\frac{\partial}{\partial z}\bigg){\cal {A}}(\bar z, z)
\vert\langle z | z' \rangle\vert^2\exp\bigg((\bar z'- \bar z).\frac{\partial}{\partial \bar z}\bigg){\cal {B}}(\bar z, z)
\end{equation}
where the overlapping of coherent states is given by
\begin{equation}
\langle z | z' \rangle = \Bigg[(1 -s \bar z'.z' )(1 -s \bar z.z )(1 -s \bar z'.z )^{-2}\bigg]^{\frac{2ks+1-s}{4}}
\end{equation}
with $s = +1 , -1$ corresponding to bosonic and fermionic statistics respectively.
Clearly, the modulus of the kernel (45) possesses the properties $\vert\langle z | z' \rangle\vert = 1$ if and only if $z=z'$, $\vert\langle z | z' \rangle\vert<1$
 and $\vert\langle z | z' \rangle\vert \to 0$ for $k$ large.
  The latter properties are helpful to get the star product between two functions
on the Bargmann space. In this respect, we introduce a function
$s(z',z)$ of the coordinates of two points on the Bargmann space
\begin{equation}
s^2(z',z) = -ln \vert \langle z | z' \rangle\vert^2 = \frac{1}{2}(2ks-s+1)ln \frac{(1 -s \bar z'.z )(1 -s \bar z.z' )}{(1 -s \bar z.z )(1 -s \bar z'.z' )}.
\end{equation}
It verifies the properties : $s(z',z) = s(z,z')$ and $s(z',z) = 0$ if and only if $z'=z$ . This
function can be interpreted as the distance between two points on the Bargmann space. It turns out that the
overlapping (45) generates the metric. In fact the line element
$ds^2$, defined as the quadratic part of the decomposition of $s^2(z, z+dz)$(distance between two infinitesimal points),
 is given by
\begin{equation}
ds^2 =  g_{i\bar j} dz_id\bar z_j
\end{equation}
where summation over repeated indices is understood and the
components of the metric $g_{i\bar j}$ are defined
\begin{equation}
g_{i\bar j} = (k + \frac{s}{2} - \frac{1}{2}) \bigg[\frac{\delta_{ij}}{1 -s \bar z.z } + s\frac{\bar z_iz_j}{(1 -s \bar z.z )^2}\bigg].
\end{equation}
We now come to the evaluation of the star product for $k$ large. Since $s^2(z',z)$ tends to infinity with $k \to \infty$ if $z\neq z'$ and equals zero
if $ z = z'$, one can conclude that, in that limit, the domain $ z \simeq z'$ gives only a contribution to the integral (44). Decomposing the intergrand near
the point $ z \simeq z'$ and going to integration over $\eta = z'-z$ , one gets
\begin{equation}
{\cal A}(\bar z, z)\star {\cal B}(\bar z, z) =  \int \frac{d\eta.d\bar{\eta}}{\pi^r} \exp\bigg(\eta.\frac{\partial}{\partial z}\bigg){\cal {A}}(\bar z, z)
\exp\bigg(-g_{i\bar j} \eta_i\bar{\eta}_{\bar j}\bigg)\exp\bigg(\bar{\eta}.\frac{\partial}{\partial \bar z}\bigg){\cal {B}}(\bar z, z).
\end{equation}
It follows that the star product between two functions on the Bargmann space associated with $A_r$ statistics is given by
\begin{equation}
{\cal A}(\bar z, z)\star {\cal B}(\bar z, z) = {\cal A}(\bar z, z)
{\cal B}(\bar z, z) - g^{i \bar j}
\frac{\partial{\cal A}}{\partial{z_i}}(\bar z, z)\frac{\partial{\cal B}}{\partial{\bar z_j}}(\bar z, z)+ O(\frac{1}{k^2})
\end{equation}
where the matrix
\begin{equation}
g^{i\bar j} =  2\frac{1 -s \bar z.z }{2k + s - 1}(\delta_{ij} -s z_i\bar z_j)
\end{equation}
is the inverse of the metric $g_{i\bar j}$ and it is proportional to $1/k$.
Then, the symbol or function associated with the commutator of two
operators $A$ and $B$ is given by
\begin{equation}
\langle z |[ A , B] | z\rangle = \{{\cal A}(\bar z, z), {\cal B}(\bar z,
z)\}_{\star} = - g^{i \bar j}\bigg(
\frac{\partial{\cal A}}{\partial{z_i}}(\bar z, z)\frac{\partial{\cal B}}{\partial{\bar z_j}}(\bar z, z)
- \frac{\partial{\cal B}}{\partial{z_i}}(\bar z, z)\frac{\partial{\cal A}}{\partial{\bar z_j}}(\bar z, z)\bigg)
\end{equation}
where
\begin{equation}
 \{{\cal A}(\bar z, z), {\cal B}(\bar z,
z)\}_{\star} = {\cal A}(\bar z, z)\star {\cal B}(\bar z, z) -
{\cal B}(\bar z, z)\star {\cal A}(\bar z, z).
\end{equation}
is the so-called the Moyal bracket.
\subsection{The excitation potential}
The quantum droplet under consideration is specified by the
density matrix $\rho_0$ (25). The excitations of this
configuration can be described by a unitary time evolution
operator $U$ which gives information concerning the dynamics of
the excitations around $\rho_0$. The excited states will be
characterized by a density operator $\rho = U \rho_0 U^{\dagger}$.
In this respect, the Hamiltonian $h$ in equation (6) may be viewed
as the excitation potential of the quantum droplet. Indeed, as we
mentioned in the previous section,   in absence of $h$ ( all $e_i$
vanishing ) the states $\vert n_1, n_2, \cdots, n_r\rangle$ are
eigenstates of $H$ with the same eigenvalue $e_0$. The degeneracy
of the energy $e_0$ coincides with the dimension of the Fock space
${\cal F}$. It is finite for $A_r$ bosonic systems and takes a
finite value for $A_r$ fermionic statistics. The Hamiltonian $h$
is exactly the excitation potential that induces a degeneracy
lifting. Using the expressions (12) and (23), the mean value of
the excitation potential $h$  is
\begin{equation}
\langle z\vert h \vert z \rangle = {\cal H}(\bar z, z) = (k +\frac{s}{2} - \frac{1}{2})\sum_{i = 1}^{r}e_i \frac{z_i\bar z_i}{1 -s \bar z.z }.
\end{equation}
The function ${\cal H}(\bar z, z)$ is the symbol associated to $h$. As we will next concerned by the edge
excitations living on the boundary of the $A_r$ quantum droplet,  it is simply verified from equations (23) and (24) that the Hamiltonian symbol
${\cal H}$ takes for $k$ large the simple form
\begin{equation}
 {\cal H}(\bar z, z) = k\sum_{i=1}^{r}e_i z_i\bar z_i,
\end{equation}
which is just the classical harmonic oscillator potential.

\section{Edge excitations and chiral bosons action}
\subsection{Effective action}
In this section, we derive the effective action for excitations living on the edge of a $A_r$ quantum droplet.
The derivation is based on semi classical analysis given in the previous
section. As mentioned above, the dynamical information, related to
degrees of freedom of the edge states, is contained in the unitary
operator $U$. The corresponding action is
\begin{equation}
S = \int dt Tr \big( \rho_0 U^{\dag}(i\partial_t - H)U \big).
\end{equation}
It is compatible with the Liouville evolution equation for the density matrix
$$ i \frac{\partial \rho}{\partial t} = [ H , \rho].$$
To write down an effective action describing the edge excitations, we evaluate
the quantities occurring in (56) as classical functions on the basis of the semi classical analysis performed above.
 We start by computing the term $ i \int dt Tr(\rho_0
U^{\dag}\partial_tU)$. For this, we set $U= e^{+i\Phi}$
$(\Phi^{\dag} = \Phi)$. A direct computation gives
\begin{equation}
dU = \sum_{n=1}^{\infty}\frac{(i)^n}{n!}\sum_{p=0}^{n-1}\Phi^p
d\Phi \Phi^{n-1-p},
\end{equation}
from which one obtains
\begin{equation}
U^{\dag}dU = i \int_0^1 d\alpha e^{-i\alpha\Phi}d\Phi
e^{+i\alpha\Phi}.
\end{equation}
Thus, we have
\begin{equation}
e^{-i\Phi}\partial_t e^{+i\Phi} = i \int_0^1 d\alpha e^{-i\alpha
\Phi}\partial_t\Phi e^{+i\alpha\Phi}.
\end{equation}
Using Baker-Campbell-Hausdorff formula, one can show
\begin{equation}
i \int dt Tr(\rho_0
U^{\dag}\partial_tU) = \int dt \sum_{n=0}^{\infty} \frac{-
(i)^n}{(n+1)!}Tr
(\underbrace{[\Phi,\cdots[\Phi}_n,\rho_0]\cdots]\partial_t\Phi)
\end{equation}
Due to the coherent states completeness, the trace of any
operator $A$ is
$$Tr A = \int d\mu(\bar z, z) \langle z | A | z \rangle.$$
It follows that the equation (60) rewrites as
\begin{equation}
i \int dt Tr(\rho_0
U^{\dag}\partial_tU) =\int d\mu dt \sum_{n=0}^{\infty} \frac{-
(i)^n}{(n+1)!}
\underbrace{\{\Phi,\cdots\{\Phi}_n,\rho_0\}_{\star}\cdots\}_{\star}\star\partial_t\Phi
\end{equation}
where the star product and the Moyal bracket are respectively defined by (50) and (52). It is
important to stress that $\rho_0$ and $\Phi$ in equation (61) are now classical functions.
It is easy to see that the equation (61) gives
\begin{equation}
i \int dt Tr(\rho_0
U^{\dag}\partial_tU) \simeq -\frac {i}{2} \int d\mu dt
\{\Phi,\rho_0\}_{\star}\partial_t\Phi
\end{equation}
where we have dropped terms in $\frac{1}{k^2}$ as well as
as the total time derivative. Using the expression (52), the Moyal bracket in (62) writes
\begin{equation}
\{\Phi , \rho_0\}_{\star} = \frac{2i}{2k+s-1}({\cal L}\Phi)  \frac{\partial\rho_0}{\partial
(\bar z.z)}
\end{equation}
where the first order differential operator is
\begin{equation}
{\cal L} =  i (1 - s\bar z.z)^2(z. \frac{\partial}{\partial z}
- \bar z.\frac{\partial}{\partial \bar z}).
\end{equation}
For $k$ large the density function (see equations (32) and (34)) is a step
function. Its derivative is a $\delta$ function with support on
the boundary $\partial {\cal D}$ of the droplet ${\cal D}$ defined
by $k\bar z.z = N$. It follows
\begin{equation}
i \int dt Tr(\rho_0
U^{\dag}\partial_tU) \approx
-\frac{1}{2}\int d\mu dt \delta(N-k\bar z.z)({\cal L}\Phi)( \partial_t\Phi )
= -\frac{1}{2} \int_{\partial {\cal
D}\times{\bf R}^+} dt ({\cal L}\Phi)( \partial_t\Phi )
\end{equation}
The second step in the derivation of edge states action consists
in the simplification  the second  term in (56) involving $H$.
By a straightforward calculation, we obtain
\begin{equation}
Tr(\rho_0 U^{\dag} h U) = Tr(\rho_0  h ) + i Tr([\rho_0, h] \Phi)
+ \frac{1}{2}Tr([\rho_0, \Phi ][h, \Phi ])
\end{equation}
The first term in r.h.s of (66) is $\Phi$-independent. We drop
it since it no contains any information about the dynamics of the
edge excitations. The second term in r.h.s of (66) rewrites as
\begin{equation}
i Tr([\rho_0, h] \Phi)  \approx i \int d\mu \{\rho_0, {\cal H}\}_{\star} \Phi
\end{equation}
in term of the Moyal bracket where ${\cal H}$ is given by (55). By
a direct computation, one can see that
\begin{equation}
i Tr([\rho_0, h] \Phi) \longrightarrow  0
\end{equation}
The last term in r.h.s of (66) gives
\begin{equation}
\frac{1}{2}Tr([\rho_0, \Phi ][h, \Phi ]) \approx  \frac{i}{2k}\int d\mu dt ({\cal L}\Phi)
\frac{\partial\rho_0}{\partial (\bar z.z)}\{{\cal H}, \Phi\}_{\star}
\end{equation}
where the Moyal bracket is given by
\begin{equation}
\{{\cal H}, \Phi\}_{\star} = \frac{2k(1-s\bar z.z)}{2k+s-1}\bigg[ s(z. \frac{\partial\Phi}{\partial z}
- \bar z.\frac{\partial\Phi}{\partial \bar z})\sum_{i=1}^r e_i\bar z_iz_i - \sum_{i=1}^r e_i (z_i \frac{\partial\Phi}{\partial z_i}
- \bar z_i\frac{\partial\Phi}{\partial \bar z_i}) \bigg]
\end{equation}
Since the derivative of the density $\rho_0$ gives a delta function with support on the boundary
of the quantum droplet, the equation (69) is simplified
\begin{equation}
\{{\cal H}, \Phi\}_{\star} \approx i \sum_{i=1}^r e_i {\cal L}_i\Phi
\end{equation}
where ${\cal L}_i$ is the angular momentum with respect to the variable $z_i$ (see the definition (64)).
Finally,  we obtain
\begin{equation}
  \int dt Tr(\rho_0
U^{\dag} H U) =  \frac{1}{2}\int d\mu dt \delta(N-k\bar z.z)({\cal L}\Phi)
\sum_{i=1}^re_i({\cal L}_i\Phi) + O(\frac{1}{k^2})
\end{equation}
Note that we have eliminated the term containing the ground states
energy $e_0$ which does not contribute to the edge dynamics.
Combining (65) and (72), we get
\begin{equation}
S \approx -\frac{1}{2}\int_{\partial {\cal D}\times {\bf R }^+} d\mu dt \delta(N-k\bar z.z)({\cal L}\Phi)
\bigg(( \partial_t\Phi )+\sum_{i=1}^r e_i ({\cal
L}_i\Phi)\bigg).
\end{equation}
This action involves only the time derivative of $\Phi$ and the
tangential derivatives $({\cal L}_i\Phi)$. It is a generalization
of a chiral abelian Wess-Zumino-Witten (WZW) theory [22]. It is
interesting to note that for $r=1$, we recover the WZW action
describing a bosonized theory of a system of large fermions in two
dimension [25]. Solving the equations of motion arising from the
action (73) gives the nature of edge states. This will be done in
the next subsection.
\subsection{Edge fields}
The action (73) is minimized by the fields $\Phi$  that satisfy the equation of motion
\begin{equation}
{\cal L}(\partial_t \Phi + \sum_{i=1}^r e_i {\cal L}_i \Phi) = 0.
\end{equation}
Since the theory is defined on the boundary of the droplet fixed
by the condition $\bar z. z = \frac{N}{k}$, we introduce the
angular variables $\theta_i $ ($ z_i = \sqrt{\frac{N}{k}}
e^{i\theta_i}$). This is the most simple parametrization that one
can consider. The operators ${\cal L}_i$ reduce to partial
derivatives $\partial_i$ with respect to $\theta_i$. The general
solution of the equation of motion can be written as
\begin{equation}
\Phi(\theta_1, \theta_2, \cdots, \theta_r, t) = \Phi(\theta_1-e_1t, \theta_2-e_2t, \cdots, \theta_r-e_rt) + \Lambda(t).
\end{equation}
In the last equation, $\Lambda(t)$ represents the gauge degree of freedom corresponding to the invariance of the action (73) under the transformation
$$ \Phi \longrightarrow \Phi + \lambda(t).$$
This can be discarded by imposing the gauge condition
\begin{equation}
(\partial_t \ + \sum_{i=1}^r e_i \partial_i )\Phi = 0.
\end{equation}
Next, we assume that the field $\Phi$ can be expressed in a factorized form
\begin{equation}
\Phi = \Phi_1 \Phi_2 \cdots \Phi_r
\end{equation}
in terms of $r$ components $\Phi_i = \Phi_i(\theta_i , t)$ ($i = 1, 2, \cdots, r$) satisfying the equations
\begin{equation}
(\partial_t \ +  e_i \partial_i )\Phi_i = 0.
\end{equation}
It is remarkable that the equations (77) and (78) are compatible with the gauge fixing condition (76). To obtain the  solution of (78), we assume
that the field $\Phi_i$ satisfies  the following periodicity condition
\begin{equation}
\Phi_i (2\pi , t) - \Phi_i (0 , t) = - 2\pi \alpha^i_0
\end{equation}
where $\alpha^i_0$ is a time independent constant. It is easy to see that the general solution of (78) is then given by
\begin{equation}
\Phi_i (\theta_i , t) =  \bar{\alpha}^i_0 - \alpha^i_0 (\theta_i - e_it) + i \sum_{n\neq 0} \frac{\alpha_n^i}{n} e^{in(\theta_i - e_it)}
\end{equation}
where the constant $\bar{\alpha}^i_0 $ can be viewed as the canonical momentum associated to $\alpha^i_0$. Note also that the complex coefficients in (80)
satisfy  $(\alpha^i_n)^{\star} =  \alpha^i_{-n}$ required by the reality condition of the field $\Phi_i$. The canonical momentum corresponding
to the field $\Phi_i$ is
\begin{equation}
\Pi_i (\theta_i , t) =   \alpha^i_0  +  \sum_{n\neq 0} \alpha_n^i e^{in(\theta_i - e_it)}.
\end{equation}
 The quantization
of the theory of edge excitations described by the action (73) can be performed by imposing the equal time commutation rules
\begin{equation}
[\Pi_i (\theta_i , t) , \Phi_j (\theta_j , t)] = i \delta_{ij} \delta(\theta_i - \theta_j).
\end{equation}
This implies that $\bar{\alpha}^i_0$ , $\alpha^i_0$ and $\alpha^i_n$ become operators satisfying the relations
\begin{equation}
[ \alpha^i_n , \alpha^j_m] = \delta_{ij} \delta_{m+n,0} {\hskip 2cm} [ \alpha^i_0 , \bar{\alpha}^j_0] = i \delta_{ij}.
\end{equation}
The other commutators vanish. This reflects that each field $\Phi_i$ is a superposition of oscillating modes on the boundary of the $A_r$ droplet.
For a fixed $i$, the Hilbert space ${\cal H}_i$ is  a tensorial product of harmonic oscillator Fock spaces. The whole Hilbert space is then given
by
\begin{equation}
{\cal H} = {\cal H}_1 \otimes {\cal H}_2 \cdots \otimes {\cal H}_r.
\end{equation}
\section{Concluding remarks}
To conclude, let summarize the main points discussed in this
paper. We provided a general approach of what we agreed to call " Chiral boson theory on Bargmann space
associated with $A_r$ quantum statistics". We started discussing the essential structures
 of these new quantum statistics.
We also gave the analytical Bargmann realizations. In the Bargmann space, the problem of
computing commutators has been rephrased
in terms of (more easy) Moyal brackets of functions associated to the algebra generated by
 creation and annihilation (Jacobson) operators. Moyal
bracket captures the essence of the full quantum characteristics of $A_r$ statistics.
This potentially provides us with an important tool to obtain
semi-classically the effective action describing a large collection of $N$ particles
 with $A_r$ statistics. More precisely, we have shown that for large $N$
and large $k$ ($k$ the parameter indexing the Fock representations), the system behaves
like a droplet in the Bargmann space (phase space). We derived the effective action
(equation (73)) describing the excitations living on the droplet's boundary. It is remarkable
 that the obtained action is similar to Wess-Zumino-action action [22]. As by product,
we have shown that the boundary excitations are essentially a
tensorial product of $r$ bosonic fields (cf. equation (77)). Each
bosonic field is given in terms of an infinite
oscillating modes (harmonic oscillators) (see equation (80)).\\
The results of the present paper can be used in connection with
the so-called generalized spin systems [27]. In fact, the usual
$SU(2)$ spin system are extended to spin models based on an
arbitrary Lie group. In particular, for the classical Lie algebras
of class $A_r$, the $SU(2)$ spin generators are replaced by the
$A_r$ generators which coincide with the creation and annihilation
(Jacobson) operators in the terminology of this paper. It follows
that the generalized spin models, discussed in [27], can be viewed
as example of systems with generalized $A_r$ statistics. In this
sense, we believe that the approach developed here can be adapted
to the generalized spin systems. We hope to report on this issue
in a forthcoming work.\\

{\bf Acknowledgments}:
One of us (M. D.) would like to thank the Max Planck Institute for Physics of Complex Systems
 for the kind hospitality extended to him during December 2007.

\end{document}